\documentclass[epsfig,natbib]{aastex}
\usepackage{emulateapj5}
\usepackage{amsmath}
\usepackage{apjfonts}

\begin{document}

\def\gtrsim{ \lower .75ex \hbox{$\sim$} \llap{\raise .27ex \hbox{$>$}} }
\def\lesssim{ \lower .75ex \hbox{$\sim$} \llap{\raise .27ex \hbox{$<$}} }

\title{ Pure and loaded fireballs in soft gamma-ray repeater giant flares}

\author{Ehud Nakar$^{1}$, Tsvi Piran$^{1,2}$ and Re'em Sari$^1$\\
{\normalsize $^1$Theoretical Astrophysics, Caltech 130-33,
Pasadena, CA, USA \\
$^2$ Racah Institute for Physics, The Hebrew University, Jerusalem 91904, ISRAEL}}

\begin{abstract}
On December 27, 2004, a giant flare  from SGR 1806$-$20 was detected
on earth. Its thermal spectrum and temperature suggest that the
flare resulted from an energy release of about $10^{47}$ erg/sec
close to the surface of a neutron star in the form of radiation
and/or pairs. This plasma expanded under its own pressure producing
a fireball and the observed gamma-rays escaped once the fireball
became optically thin. The giant flare was followed by a bright
radio afterglow, with an observable extended size, implying an
energetic relativistic outflow. We revisit here the evolution of
relativistic fireballs and we calculate the Lorentz factor and
energy remaining in relativistic outflow once the radiation escapes.
We show that  pairs that arise naturally in a pure pairs-radiation
fireball do not carry enough energy to account for the observed
afterglow. We consider various alternatives and we show that if the
relativistic outflow that causes the afterglow is related directly
to the prompt flare, then the initial fireball must be loaded by
baryons or Poynting flux. While we focus on parameters applicable to
the giant flare and the radio afterglow of SGR 1806$-$20 the
calculations presented here might be also applicable to GRBs.
\end{abstract}

\maketitle

\section{Introduction}

The giant flare from SGR\,1806$-$20  was the most powerful flare of
$gamma$-rays ever measured on earth. It lasted about $0.2$sec. Its
fluence of  $\approx 1\,$erg cm$^{-2}$ \citep{hbs05,pbg05},
correspond to energy of $3 \times 10^{46}\,$erg released at a
distance of\footnote{Through out  the paper we consider a distance
of 15kpc even though SGR 1806-20 might be closer by a factor of
$\lesssim 2$ (\cite{fnk04,ngp05,cck05} see however \cite{mg05}). We
express the dependence of our calculations on the flare energy as
implied from this distance}. Our conclusions do not depend on the
exact distance. 15kpc \citep{ce04}. This energy exceeds the energies
of the giant flares from SGR 0526$-$66 (the famous March 5$^{th}$
1979 event) and from SGR 1900+14 (August 27$^{th}$ 1998) by a factor
of a hundred \citep{m99}. The spectrum of the flare is consistent
with that of a cooling blackbody spectrum with an average
temperature of $175 \pm 25$keV \citep{hbs05}. The reflection from
the moon, as detected by Helicon-Corona-F \citep{gcn2923} with a
fluence of $7.5 \times 10^{-7}$erg cm$^{-2}$, provides an
alternative mean of estimating the fluence at the energy range
25-400keV. The albedo of the moon in this energy range is around
0.25 \citep{ngp05}, resulting in an isotropic energy release of
about $10^{46}\,$erg. This value is consistent with the fluence and
the spectrum measured by RHESSI, since for a 170~keV black body,
only a quarter of the energy is radiated in the $25-400$~keV range.

Like the two other giant flares this flare was followed by a pulsed
softer X-ray emission that lasted more than 380sec \citep{gcn2922}.
Radio afterglow was detected from VLA observations
\citep{cck05,gkg05}\footnote{A radio afterglow was detected also
after the August giant flare from SGR 1900+14 \citep{fkb99}}. After
one week the radio source was extended with a size of $\theta=40-80
$masec corresponding to a radius of $0.6-0.9 \times 10^{16}$cm and
an average velocity of 0.3c-0.5c (at $d=15$kpc). Therefore a
significant amount of energy was emitted in the form of a
relativistic ejecta around the same time that the $\gamma$-rays
flare was emitted.

According to the standard magnetar model
(\citealt{dt92,p92,k94_22,td95,td96,wkg+01,e02}; see also
\citealt{k82} for related early ideas) , the giant flare is produced
via annihilation of the magnetic field of a highly magnetized
neutron star. This annihilation deposits energy at the form of
photons and pairs near the surface of the neutron star.  The
pair-radiation plasma evolves as an accelerating fireball
\citep{g86,p86,sp90,psn93,mlr93,gw98} resulting in a thermal
radiation burst carrying the bulk of the initial energy with roughly
the original temperature and a fraction of the energy in the form of
relativistic pairs. The thermal spectrum of the flare and its
temperature support this picture.

Here we compare the energy required to produce the radio afterglow
(calculated in \S 2) with a new simple calculation (\S 3) of the
energy of the pairs outflow. As the available pairs energy is short
by at least two orders of magnitude we consider (\S 4) baryonic or
electromagnetically loaded fireballs, again providing new simple
estimates for these cases. We compare our calculations of the
fireball evolution to previous works in \S 5 and find that our
simple estimates correct previous works. Finally, we discuss the
implications of our results to the giant flare of Dec 27$^{th}$
flare from SGR\,1806$-$20 and of August 27$^{th}$ from
SGR\,1900$+$14 in \S 6.

\section{A lower limit on the radio afterglow's energy}

The energy emitted in the radio during the second week after the
burst is about $10^{38}$erg \citep{cck05,gkg05}. This is clearly an
absolute lower limit to the energy of the relativistic ejecta. A
stronger limit can be obtained if we assume synchrotron emission as
suggested by the optically thin spectrum, $F_\nu \propto \nu^{-0.7}$
and the observed linear polarization \citep{gkg05}. We employ here
the familiar equipartition method \citep{p70,sr77}. To emphasize the
robustness of this method, which depends only on the assumption of
synchrotron radiation, we sketch the derivation here. We
characterize the emitting region by the number of its pairs $N$, the
magnetic field, $B$, and the typical thermal Lorentz factor of the
pairs, $\gamma_e$. We ignore the mildly relativistic motion. The
most conservative assumption that all the electrons emit at a
synchrotron frequency of $\nu_R=8.5$GHz requires that the pairs have
a Lorentz factor of:
\begin{equation}\label{Eq gamma_e}
\gamma_e \approx \left(\frac{2\pi m_ec \nu_R}{e B}\right)^{1/2}.
\end{equation}
While to obtain the observed flux the number of pairs must satisfy:
\begin{equation}\label{Eq FnuR}
   N   \approx  \frac{12\pi d^2 e F_{\nu,R}}{\sigma_T m_e c^2 B}.
\end{equation}
The sum of the energies of the pairs and of the magnetic field,
$E=R^3B^2/6 + N m_ec^2\gamma_e$, is minimized once the former is 3/4
of the latter:
\begin{equation}\label{EQ Emin synch}
 \begin{array}{c}
                  E_{min} \approx (5 \times 10^{42} ~{\rm ergs}~)
\left(\frac{\theta}{0.065''}\right)^\frac{9}{7}\left(\frac{d}{15kpc}\right)^\frac{17}{7} \\
                  \times\left(\frac{F_{\nu,R}}{50mJy}\right)^\frac{4}{7}\left(\frac{\nu_R}{8.5GHz}\right)^\frac{2}{7}. \\
                \end{array}
\end{equation}
We assumed here, conservatively, that all the emission is radiated
at $\nu_R$. As the observed afterglow shows a spectral power-law
over more than an order of magnitude in frequency our estimate for
$E_{min}$ should be larger by a factor of  5
\citep{p70,cck05,gkg05} bringing it to $\sim 3 \times 10^{43}$
erg.

If the radio afterglow of this SGR arises, like in GRBs or
SNRs, from shocks going into the surrounding medium, farther
constraint exists. The observed shock size determines its velocity
and therefore the typical electron Lorentz factor. The external
density and the Lorentz factor dictate the magnetic field. These
additional constraints will force $\gamma_e$ and $N$ to deviate from
equations \ref{Eq gamma_e} \& \ref{Eq FnuR} and therefore result in
a higher energy estimate.

\section{Pairs-Radiation Fireball}
Consider an energy $E = 10^{46}$ erg that is deposited in the vicinity
of a neutron star ($R_0 \approx 10^6$cm) as photons with a typical
energy $ e_\gamma\approx 500\,$keV (with a black body temperature of
$170$~keV, more than half of the energy is in photons with
$e>e_\gamma$). If the energy is released instantaneously then the
duration of the observed emission would be $R_0/c \approx 0.1{\rm ms}
R_{0,6}$. To be compatible with the observed duration of the flare,
the source must have been active for a time comparable to the observed
duration, $t \approx 0.1$ sec. The optical depth for pair production
would be:
\begin{equation}\label{EQ tau}
    \tau_{\gamma \gamma} \gtrsim \frac{E \sigma_T}{4\pi e_\gamma
    R_0 c t} \approx 2 \times 10^{11} {E_{46} \over R_{0,6} t_{-1} } ,
\end{equation}
where $\sigma_T$ is the Thompson cross-section and $N_x$ denotes
$N/10^{x}$ in cgs units. The large magnetic fields of the magnetar,
decrease the effective cross-section \citep{H79} but not
sufficiently to make this radiation optically thin. With such a
large optical depth the radiation forms a radiation-pairs plasma at
a thermal equilibrium with an initial temperature of
\begin{equation}\label{EQ temp}
    T_0 \approx  \left( \frac{E}{4\pi R_0^2\sigma t}\right)^{1/4}
    \approx 300~ {\rm keV}~ E_{46}^{1/4} R_{0,6}^{-1/2}t_{-1
    }^{-1/4},
\end{equation}
where $\sigma$ is the Stephan-Boltzmann constant. This
radiation-pairs plasma expands relativistically
\citep{g86,p86,sp90,psn93,mlr93,k96,td96}  with $\Gamma \propto R$
and $T \propto R^{-1}$ until at $R_\pm$ the pairs stop annihilating
and their number freezes. The number of remaining pairs, $N_{\pm}$,
is determined by the condition that their annihilation rate $
n_{\pm} \sigma (\beta_\pm) \beta_\pm c $ equals the local expansion
rate $c/(R/\gamma) \approx c/R_0$:
\begin{equation}\label{Eq Npm}
N_{\pm} ={4\pi R_0ct T_0^3  \over \sigma_T T_\pm^3} \approx 2.7
\times 10^{44} E_{46}^{3/4}R_{0,6}^{-1/2}t_{-1}^{1/4} ,
\end{equation}
where $n_{\pm}$ is the pairs density and $T_\pm$, the temperature in
the fireball rest frame at $R_\pm$, is $\sim 18$keV
\citep{g86,p86,sp90}. We have used here the fact that at low energy
the cross section for annihilation is $\sigma(\beta_{\pm}) \approx
\sigma_T /\beta_\pm$ with $\beta_\pm c =v_\pm$ the thermal velocity
of the pairs in the local frame. The photons decouple from the pairs
around the same time that the pairs freeze out. While at this stage
the thermal velocity of the pairs $v_\pm$ is much smaller than $c$,
the cross section for annihilation is larger than the cross section
for photon scattering by the same factor.  The escaping photons,
which carry most of the energy, have a quasi-thermal spectrum
\citep{g86,gw98} with $T_{obs} = \Gamma T_\pm = T_0$, roughly the
initial temperature.

Even after the pairs stop annihilating and the photons decouple from
the pair the pairs do not decouple from the photons.  The huge
photon flux continues to accelerate the pairs \citep{mlr93,gw98}.
The acceleration continues as long as the force that the photon
field applies on an electron (positron) is sufficient to accelerate
the electron so that it remains at the same Lorentz factor as the
bulk of the photon field. The condition for effective acceleration
is that during the time that the radius doubles the work done by the
photon field on an electron $E \sigma_T/4 \pi R ct\gamma^2$ is
larger than $\gamma m_e c^2$, the energy that the electron needs to
gain during this period in order to keep up with the accelerating
flow. This implies that the pairs accelerate until they reach a bulk
Lorentz factor of :
\begin{equation}
\Gamma_{\pm} =   \left(E \sigma_T \over  4 \pi c^3 t m_e
R_0\right)^{1/4} \approx 680 E_{46}^{1/4}R_{0,6}^{-1/4}t_{-1}^{-1/4}
, \label{gamma_max_pairs}
\end{equation}
and their kinetic energy is \footnote{This result corrects the one
presented in \cite{gw98} that find no dependence of the final
pairs energy on the initial radius. We show here that $E_{\pm}
\propto R_0^{-3/4}$. The value of $E_{\pm}$ for $R_0=10^6$ cm is
similar in both works (see \S\ref{Sec comparison}).}:
\begin{equation}
E_{\pm}= N_\pm m_ec^2\Gamma_\pm \approx 1.4 \times 10^{41} {\rm
erg}~E_{46} R_{0,6}^{-3/4}.
\label{Epm}
\end{equation}
This kinetic energy is smaller by two orders of magnitude than the
minimal energy required to produce the observed radio flux (Eq.
\ref{EQ Emin synch}). Therefore, the energy source of the radio
afterglow cannot be the kinetic energy that remain in the pairs
outflow.

It is important to note that while strong magnetic fields, that might
be dragged from the magnetar into the fireball, may influence the
interaction between the photons and the pairs they do not change the
conclusion.  Strong magnetic fields ($B \gtrsim 10^{13}G$) would
decrease the cross-section for photon-electron (positron) scattering
\citep{H79} resulting in a smaller $\Gamma_{\pm}$. In addition, a
strong magnetic field would suppress the cross-section for pair
annihilation into two photons, but it would also open a new channel of
pair annihilation into a single photon \citep{W79,H86}. The latter
becomes the dominant annihilation process and its cross-section is
larger than the cross-section with no magnetic field that was used in
Eq. \ref{Eq Npm}. The overall result is a lower $N_\pm$. Thus, strong
magnetic field would only decrease the energy that remain in the
pairs.  The energy of the magnetic field itself may, however,
contribute a significant component to the energy of the relativistic
outflow. We address this contribution in section \S\ref{loaded}.

\subsection{Interaction between the radiation and the circum burst
medium}

The energy needed to power the afterglow  is only a small fraction
of the total prompt $\gamma$-ray energy.  It is therefore worthwhile
exploring whether the interaction of the prompt radiation with the
circum-flare medium can give rise to a relativistic outflow with the
required energy. The optical depth is given by $\sigma_T n R$, where
$n$ is the ambient density and $R \approx 10^{16}$~cm is the
observed size of the radio emitting region. The energy acquired by
the ambient electrons within $R$ is therefore
\begin{equation}
{E_{r<R}\over E}=\sigma_T n R=5\times 10^{-9} n R_{16}
\end{equation}
Tapping $10^{-3}$ of the energy requires an unreasonably large
average density of $n \ge 10^5 {\rm cm^{-3}}$. This is not
expected around the magnetar given possible previous bursts.
Moreover, giving this amount of energy to such a large mass would
result in a sub-relativistic velocity ($v \ll 0.1c$).

An alternative mechanism can be pair enrichment by collisions of the
outgoing radiation with photons that are back-scattered by the
ambient medium \citep{TM00,B02,MRR01}. These pairs would in turn
scatter more photons that would create more pairs and so forth. This
process will take place up to a radius where each ambient electron
scatters at least one photon:
\begin{equation}\label{EQ Rpairs}
    R_{enrich} = \left(\frac {E\sigma_T}{4\pi
    \epsilon_\gamma}\right)^{1/2} = 2.5 \times 10^{13} {\rm cm}
    E_{46}^{1/2}.
\end{equation}
At smaller radii the number of pairs will grow exponentially until
there will be about $m_p/m_e$ pairs for each ambient medium
electron. At this point there are enough pairs to accelerate the
ambient medium to a relativistic velocity. Once this happen the
temperature of the radiation in the rest frame of the accelerated
medium drops significantly. Since the spectrum of the radiation is
thermal there are no photons with an energy larger than $m_ec^2$ at
this frame. The scattered photons stop creating additional pairs and
the exponential process stops. Thus, up to $R_{pairs}$ the density
of pairs is expected to be about $m_p/m_e$ times larger than the
external medium density. The fraction of energy acquired by this
pair enriched region is
\begin{equation}
{E_{enrich} \over E} = {m_p \over m_e} \sigma_T R_{enrich} n\cong
10^{-8} E_{46}^{1/2} n_0
\end{equation}
 Again, this enrichment is insufficient to tap $10^{-3}$ of the
initial energy unless the external medium average density is $\sim
10^5 {\rm cm^{-3}}$. Even if one sets such large density around the
neutron star, the high density region must be truncated shortly
after $3 \times 10^{13}$cm, to allow for the observed mildly
relativistic motion. Such a configuration seems to be too contrived.

We conclude that the interaction between the prompt radiation and the
external medium is unlikely to be the source of the afterglow energy.

\section{Loaded Fireballs \label{loaded}}

\subsection{Baryonic Load}

The processes that govern the evolution of a fireball loaded with
protons are similar to those that govern a pure pairs-radiation
fireball.  However, the electrons that accompany the protons
contribute to the opacity while the protons contribute to the
inertia. Both effects can be taken into account by generalizing Eqs.
\ref{gamma_max_pairs}-\ref{Epm}. To do so we replace $m_e$ with the
mean mass per particle:
\begin{equation}\label{Eq m_avg}
    \overline{m}=m_p \frac{(m_e/m_p)N_\pm  + N_p}{N_\pm + N_p},
\end{equation}
and we replace $N\pm$ with the total electrons and positrons
density:
\begin{equation} N=N_p+N_\pm ,
\end{equation}
where $N_p=E/(m_pc^2\eta)$ is the number of protons and  $\eta
\equiv E/M c^2$ characterize the baryonic load. The generalized
equations are valid as long as the baryons load is small enough and
the radiation escapes before most of its energy is converted to the
baryonic kinetic energy.

The behavior of $\overline{m}$ and $N_{B\pm}$ as a function of the
baryon load, $\eta$, depends on the ratio $N_p/N_\pm$. There are two
critical values of this ratio: $N_p/N_\pm = m_e/m_p$ and $N_p/N_\pm
= 1$. The former marks equal mass for baryons and pairs, the latter
marks equal contribution to the Thompson scattering.  These values
corresponds respectively to the following critical values of $\eta$:
\begin{equation}\label{Eq eta_cr1}
    \eta_{1}={E \sigma_T \over 4\pi R_0 ct m_e c^2}\left( T_\pm \over T_0 \right)^3=4.5 \times 10^7 E_{46}^{1/4}R_{0,6}^{1/2}t_{-1}^{-1/4}
\end{equation}
\begin{equation}\label{Eq eta_cr2}
    \eta_{2}={E \sigma_T \over 4\pi R_0 ct m_p c^2}\left( T_\pm \over T_0 \right)^3=2.5 \times 10^4 E_{46}^{1/4}R_{0,6}^{1/2}t_{-1}^{-1/4}
\end{equation}
An additional critical value of $\eta$ is defined by the condition
that the photons have transferred effectively all their energy to the
baryons:
\begin{equation}\label{Eq eta_cr3}
    \eta_{3}= \left( E \sigma_T \over 4\pi R_0 c t m_p c^2 \right)^{1/4}=100 E_{46}^{1/4}R_{0,6}^{-1/4}t_{-1}^{-1/4}.
\end{equation}
For this critical $\eta$ the photons decouple from the baryons just
at the moment that the baryons stop being accelerated by the
photons.

Figure 1 illustrates the dependance of the energy that remain in the
plasma, $E_{p,\pm}$ and its final Lorentz factor $\Gamma_{p,\pm}$ on
$\eta$. When $\eta_{1} \ll \eta$ the baryons do not affect the
evolution, $\overline{m} \approx m_e$ and $N \approx N_\pm$ and
therefore the fireball evolves as a pure pairs-plasma fireball.  The
bulk Lorentz factor and energy are given by equations
\ref{gamma_max_pairs} and \ref{Epm}. When $ \eta_{2} \ll \eta \ll
\eta_{1}$ the baryons carry most of the inertia while the pairs are
still responsible for most of the opacity. Therefore, $\overline{m}
\approx m_e \eta_{1}/\eta$ while $N \approx N_\pm$ and
 $E_{p,\pm} \propto \eta^{-3/4}$
while its final Lorentz factor $\Gamma_{p,\pm} \propto \eta ^{1/4}$.
In this case the pairs become transparent for the radiation at
$R_{\pm}$. For $ \eta_{3} \ll \eta \ll \eta_{2}$ the baryons provide
both the inertia and the opacity. The final Lorentz factor is
constant and $E_{p,\pm} \propto \eta^{-1}$. The radiation decouple
from the electrons at radius larger than $R_\pm$, but the radiation
still carry most of the energy and its temperature is $\approx T_0$.
For $\eta \ll \eta_{3}$ the radiation decouples from the baryons
only after it transferred most of its energy to the baryons. The
baryons energy is the total energy $E$ and their Lorentz factor is
$\eta$. The radiation energy however is much smaller than $E$ and
its temperature is much smaller than $T_0$.  We call this ($\eta <
\eta_{3}$) a heavy load.

The final Lorentz factor can be approximated as (see Fig 1):
\begin{equation}\label{EQ GammaBpm}
    \Gamma_{p\pm} \approx \left \{ \begin{array}{lc}
                            680 E_{46}^{1/4}R_{0,6}^{-1/4}t_{-1}^{-1/4} &  \eta_{1} < \eta  \\
                            680 (\eta/\eta_{1})^{1/4} E_{46}^{1/4}R_{0,6}^{-1/4}t_{-1}^{-1/4} & \eta_{2} < \eta < \eta_{1} \\
                            100 E_{46}^{1/4}R_{0,6}^{-1/4}t_{-1}^{-1/4} & \eta_{3} < \eta < \eta_{2} \\
                            \eta & \eta < \eta_{3} \\
                         \end{array} \right..
\end{equation}
While the energy that remain in the ejecta is:
\begin{equation}\label{EQ EBpm}
   \frac{E_{p\pm}}{E} \approx \left \{ \begin{array}{lc}
                             1.4 \times 10^{-5} R_{0,6}^{-3/4}&  \eta_{1} < \eta  \\
                             4 \times 10^{-3}  (\eta/\eta_{2})^{- 3/4} R_{0,6}^{-3/4}  & \eta_{2} < \eta < \eta_{1} \\
                             \eta_{3}/\eta & \eta_{3} < \eta < \eta_{2} \\
                             1  & \eta < \eta_{3} \\
                         \end{array} \right..
\end{equation}

So far we have only considered protonic loading. The evolution of a
neutron rich fireball was explored in the past by several authors
\citep{dkk99,b03,rbr04,vpk03}.  The neutrons are coupled to the
plasma only through collisions with protons. Initially the protons
drag the neutrons efficiently.  A neutron that collides with a
proton receives in such a collision at most 1GeV. Therefore, the
dragging takes place as long as each neutron collides with at least
one proton during the time that the protons double their Lorentz
factor. This happens when:
\begin{equation}
R n_p \sigma_0/\Gamma  \ge 1
\end{equation}
where $n_p$ is the protons density and $\sigma_{0} =
\sigma_{np}\beta_{rel} \approx 3 \cdot 10^{-26}{\rm cm}^2$ where
$\sigma_{np}$ is the neutron-proton cross section and $\beta_{rel}$
is the relative velocity between neutrons and protons.

With $\sigma_{0}$ about an order of magnitude lower than $\sigma_T$,
the neutrons always decouple from the protons before the radiation
decouples from the plasma. If the neutrons do not decouple before
the acceleration ends the load is essentially heavy and the photons
transfer most of their energy to the baryons. The final Lorentz
factor in this case is $\eta$, corresponding to the total baryonic
load. If the neutrons decouple before the acceleration ends then
during the decoupling phase (when the neutrons begin to lag after
the protons) the neutrons become relativistically hot and the
inelastic n-n collisions convert neutrons to protons while producing
pions.  If initially $N_n/N_p \gg 1 $, a significant fraction of the
neutrons will be converted to protons and this ratio will become of
order unity \citep{dkk99} or somewhat larger \citep{fpa00} after
decoupling. Thus the final protonic load, that carries most of the
energy, is comparable to the initial total baryonic load.  The final
Lorentz factor and the energy of the plasma can be approximated by
Eqs. \ref{EQ GammaBpm} \& \ref{EQ EBpm} if $\eta$ is defined
according to the total baryonic load.

\subsection{Heavy Loading}
For completeness we also consider here the energy and the
temperature of the thermal radiation that escapes from a heavily
loaded fireball, namely a fireball with $\eta < \eta_{3}$. This
situation is clearly inapplicable to this giant flare, but it may be
relevant to GRBs, possibly as an explanation of precursors which are
less energetic than the burst itself. In this case the Lorentz
factor saturates at $R_\eta=R_0\eta$, and the fireball enters its
coasting phase. The fireball is coasting without spreading (both in
the local and the observer frame) until $R_{spread}=ct\eta^2$. At
larger radii the fireball spreads to a width of $\sim R/\eta^2$ in
the observer frame. The radiation in this case ($\eta < \eta_{3}$)
decouples from the matter long after $R_\eta$ at:
\begin{equation}
R_{ph}=\left\{ \begin{array}{c}
          {E \sigma_T \over 4\pi c t m_p c^2 \eta^3} ~~~ R_{ph}<R_{spread} \\
          \left({E \sigma_T \over 4\pi m_p c^2 \eta}\right)^{1/2} ~~~ R_{ph}>R_{spread} \\
        \end{array} \right..
\end{equation}

During the acceleration phase ($R<R_\eta$) the photons cool in their
rest frame.  The acceleration compensates for this cooling keeping
the temperature and the overall radiation energy in the observer
frame, $E_{ph}$, constant. During the coasting phase, with no
acceleration, the photons cool and lose energy in the observer frame
as well. Since the ratio of the photon number to the proton number,
$\sim m_pc^2\eta/T_0 \gg 1$, is roughly constant at all times, the
photons govern the cooling with an adiabatic index of 4/3
($TV^{1/3}=const$, where V is the volume). For
$R_0\eta<R<\min(R_{ph},R_{spread})$, i.e during the coasting phase
while the fireball is still opaque, $V \propto R^2$ and therefore
$T\propto R^{-2/3}$ and $E_{ph} \propto R^{-2/3}$. If the fireball
is still opaque at $R_{spread}$ then for $R_{spread}<R<R_{ph}$ the
radiation evolves with $T\propto R^{-1}$ and $E_{ph} \propto
R^{-1}$. Thus as far as the radiation is concerned there is another
critical $\eta$ for which $R_{ph}=R_{spread}$:
\begin{equation}\label{EQ eta4}
    \eta_4=8E_{46}^{1/5}t_{-1}^{-2/5}
\end{equation}
For $\eta_4<\eta<\eta_3$:
\begin{equation}\label{EQ ET_eta34}
    \begin{array}{c}
      \frac{R_{ph}}{R_\eta}=\left(\frac{\eta_3}{\eta}\right)^4, \\
      \\
      \frac{E_{ph}}{E}=\frac{T_{obs}}{T_0}=\left(\frac{\eta_3}{\eta}\right)^{-8/3}= \left(\frac{\eta}{100}\right)^{8/3}E_{46}^{-2/3}R_{0,6}^{2/3}t_{-1}^{2/3} \\
    \end{array}
\end{equation}
For $1<\eta<\eta_4$:
\begin{equation}
    \begin{array}{c}
      \frac{R_{spread}}{R_\eta}=\frac{ct\eta}{R_0}=\left(\frac{\eta_3}{\eta_4}\right)^4\frac{\eta}{\eta_4}~~~;~~~\frac{R_{ph}}{R_{spread}}=\left(\frac{\eta_4}{\eta}\right)^{5/2} , \\
      \\
      \frac{E_{ph}}{E}=\frac{T_{obs}}{T_0}=3 \times 10^{-5} E_{46}^{-1/2}R_{0,6}^{2/3}t_{-1}^{1/3}\eta^{11/6} \\
    \end{array}
\end{equation}


In GRBs the final Lorentz factor of the relativistic ejecta is
$\gtrsim 300$, as indicated by opacity considerations \citep{ls01},
and therefore $\eta \gtrsim 300$. On the other hand, the non-thermal
spectrum of the prompt emission implies $\eta \lesssim \eta_3 \sim
10^3$ for GRBs, making the range of allowed $\eta$ very narrow. It
implies also that the ratio between the energy in the thermal
radiation that escapes the fireball and its kinetic energy is
$(\eta_3/\eta)^{-8/3}\gtrsim (1000/300)^{-8/3} \approx 5\%$. Taking
into account the efficiency in which this kinetic energy is
converted into a non-thermal radiation, it indicates that there
should be a non-negligible thermal component in almost any GRB. This
result may be supported by the observations, as suggested by
\cite{r05}. The idea that a thermal component in the prompt emission
is indeed the radiation that escapes from the fireball can be tested
simply by comparing the energy and the temperature in this
component:
\begin{equation}
    T_{obs} = 1MeV E_{th,51} E_{52}^{-3/4}
    R_{0,6}^{-1/2}t_{-1}^{-1/4}.
\end{equation}
Where $E_{th}$ is the energy in the thermal component and $E$ is the
total observed energy (note that we use here the typical values for
GRBs). This relation should be tested within a given burst pulse by
pulse.

\begin{figure*}
\epsscale{1.4} \plotone{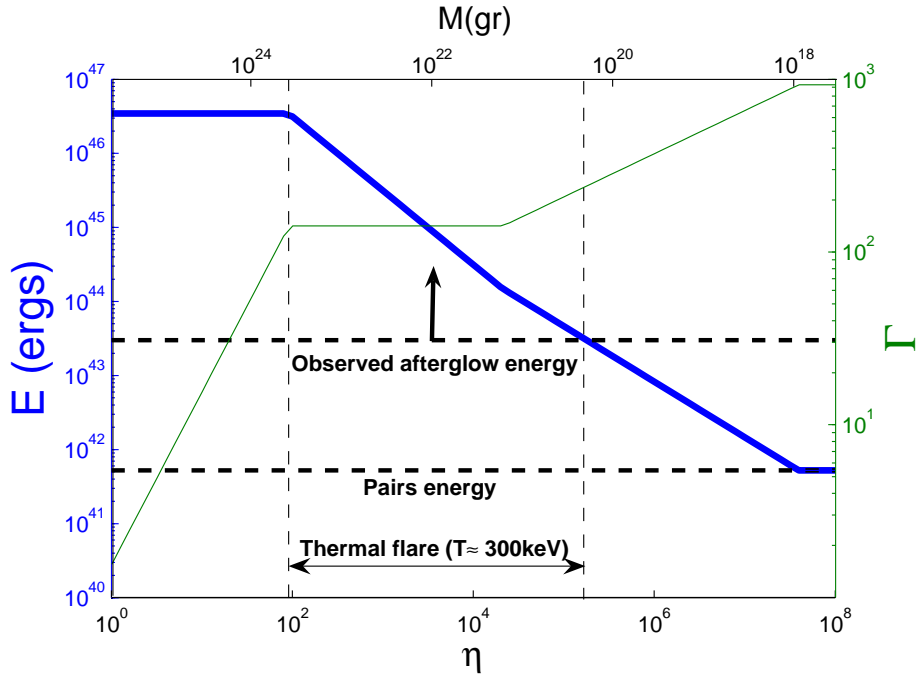}

\caption{The baryons' energy (thick line, left y-axis) and the
baryons' final Lorentz factor (thin line, right y-axis) as functions
of $\eta$ (lower x-axis). For clarity the total baryons mass is
depicted on the upper x-axis. Also marked (short dashed horizontal
lines) are the lower limit on the energy inferred from the radio
afterglow and the energy that remains in the pairs in a pure
radiation-pairs fireball. The vertical dashed lines mark the allowed
loading in order to produce a thermal flare at the observed
temperature. The parameters in this plot are  $E=3.5 \times 10^{46}$
erg, $t=0.1$sec and $R_0=10^6$ cm.}

\end{figure*}

\subsection{Electromagnetic load}

An alternative loading of the fireball is an electromagnetic load.
Consider a magnetic field that is confined to the fireball and that
is accelerated with it. The magnetic field carries an energy $\gamma
E_{EM}$ where $E_{EM}$ is the electromagnetic energy in the
fireball's rest frame at the time that the pairs decouple from the
radiation. The electromagnetic loading can be treated similarly to
the baryonic loading where the magnetic field contribution to the
inertial mass is:
\begin{equation}\label{EQ m avg EM}
    \overline{m}=m_e+\frac{E}{\eta_{EM}N_\pm c^2},
\end{equation}
and $\eta_{EM}=E/E_{EM}$ is  a measure of the electromagnetic
loading. The total number of pairs, and therefore the opacity, is
similar to those of a pure pairs-radiation fireball \footnote{ This
is true as long as the magnetic field is much smaller than
$10^{13}$G at $R_\pm$, so the cross-section for Compton scattering
is not affected. In our case where $E \approx 10^{46}$ erg, $R_\pm
\approx 10^7$cm and if the relevant loading is $\eta_{EM} \gtrsim
10$ this condition is satisfied.}:
\begin{equation}
N=N_\pm .
\end{equation}
Therefore, there are only two critical values of $\eta_{EM}$:
$\eta_{EM,1}=\eta_{1}$ and $\eta_{EM,2} = T_0/T\pm$. When $\eta_{1}
\ll \eta_{EM}$ the fireball evolves as a pure pairs-plasma fireball.
When $ \eta_{EM,2} \ll \eta_{EM} \ll \eta_{EM,1}$ the radiation
still carries most of the energy and its observed temperature is
$T_0$. The total energy that is left in the magnetized pairs after
they decouple from the radiation is $E_{EM,\pm} \propto
\eta^{-3/4}$. Finally, when $\eta_{EM} \ll \eta_{EM,2}$ the
radiation decouples from the magnetized pairs only after it has
transferred to them most of the energy. In this case $E_{EM,\pm} =
E$ while The radiation energy is $E_{ph}=E(\eta_{EM,2}/\eta)^{-8/3}$
and its temperature $T_{obs}=T_0(\eta_{EM,2}/\eta)^{-8/3}$. Note
that when the electromagnetic loading is significant ($
\eta_{EM}<\eta_{1} $) the Lorentz factor of the pairs at the time
that they decouple from the radiation is only a lower bound on their
final Lorentz factor. The reason is that energy transfer between the
electromagnetic field and the pairs can still take place at larger
radii.

\section{Comparison with previous works}\label{Sec comparison}
The evolution of pure and loaded fireballs was explored by several
authors in the past. \cite{p86} and \cite{g86} considered only the
properties of the radiation that emerges from a pure fireball.
\cite{sp90}, \cite{psn93} and \cite{mlr93} considered the evolution
of a loaded fireball. These papers miscalculated the radius that the
baryons decouple from the radiation, resulting in an overestimate of
$\eta_3$ and of the final Lorentz factor and the ejecta energy,  for
$\eta > \eta_{3}$. In the context of GRBs our result ($\eta_3 \sim
10^3$) is significantly lower than in those papers, and combined
with the lower limits on the Lorentz factor, $\eta \approx 500$ is
narrowly constrained. More recently, \cite{gw98} carried out a
detailed numerical and analytical calculation of the final Lorentz
factor and energy of pure and loaded fireballs. Our much simpler
estimates generally agree with their results with the exception of
the final energy that remain in the pairs, $E_{\pm}$ and their
opacity for the radiation once the Lorentz factor saturates
$\tau_{\pm}$.  While \cite{gw98} find that both quantities do not
depend on the initial radius, we find that $E_{\pm} \propto
R_0^{-3/4}$ and that $\tau_{\pm} \propto R_0^{-3/4}$ as well.
However, the value that these authors obtain for $R_0=10^6$ cm are
similar to the values that we present here.

\section{Discussion}

We presented here a simple, yet comprehensive, derivation of the
evolution of pure and loaded fireballs. Our results correct
previous works (see \S\ref{Sec comparison}), enabling us to
compare the energy that remains in the plasma once the radiation
escapes to the observations of the radio afterglow of the giant
flare from SGR 1806-20.

The relativistic ejecta that produced this radio afterglow
contained at least $0.1\%$ of the energy emitted in $\gamma$-rays.
The thermal spectrum of the flare and its temperature indicate
that the flare resulted from an energy deposition near the neutron
star surface. Such energy deposition must create an opaque
accelerating fireball. An elegant explanation for the afterglow
energy source could have been the inevitable energy of the
remaining pairs after they decouple from the radiation. However,
we find that this energy is short by at least two orders of
magnitude, excluding this possibility. We considered also the
possibility that the energy of the afterglow is obtained by the
interaction of the outgoing radiation with the external medium. We
find this scenario to be unlikely since it requires a rather high
external density with a contrived profile.

We conclude, therefore, that if the relativistic ejecta that
produces the afterglow is directly related to the prompt flare
(rather than e.g. to the following confined fireball), then the
fireball must have been loaded with either baryons or Magnetic field
in order to enhance the energy that remains in this ejecta. This
loading should be however fine tuned in order to obtain the right
amount of energy. Even if the afterglow energy is comparable to that
of the ejecta (we have only lower limit on this energy) the loading
cannot be too high. If $\eta < \eta_{3}$ the energy in the flare
drops significantly and so does the temperature. The agreement of
the observed temperature and the black body temperature estimated in
Eq. \ref{EQ temp} suggest that this is not the case. For the
December $27^{th}$ giant flare the range of the allowed baryonic
loading is $100 \lesssim \eta \lesssim 10^5$, which corresponds to a
baryons mass of $10^{20} {\rm gr} \lesssim m_b \lesssim 10^{23} {\rm
gr}$ (see fig. 1).

Interestingly similar lower limit on the ratio of the afterglow
energy to the flare energy was observed in the August giant flare
from SGR 1900+14 \citep{fkb99}, although this flare was hundred
times dimmer than the December $27^{th}$ giant flare.  With the
caveats of small number statistics, and both afterglow estimates
being lower limit, this similarity suggests a common origin for the
two afterglows and a linear relation between the observed afterglow
energy and the flare energy. The similarity between the fractional
energy in the afterglows of the two events, is not accounted for in
our model. The most natural energy source, the pairs energy in a
baryon free fireball, contain a constant fraction of the fireball
energy (see Eq.  \ref{Epm}) but this fraction is far too short. If
this similarity is confirmed by future events it may indicate that
some process regulates the baryonic load. For example it is possible
that a constant number of baryons per unit energy are torn apart
from the surface of the neutron star and are mixed with the
fireball.  Alternatively it is possible that a fixed fraction of the
energy is ejected as a Poynting flux.  A different solution might be
that the fireball is  heavily loaded aspherically, such that only a
small portion of the $4 \pi$ solid angle is loaded, while the rest
of the fireball is pure. In this case $E_{p,\pm}/E$ is simply the
ratio of loaded portion solid angle to the pure portion solid angle,
and this ratio might be similar between different flares. It is, of
course, also possible that the afterglow is not related directly to
the prompt flare and that it is created by an independent mechanism
and its energy is not extracted of the initial fireball. Once more
the similarity between the ratios of afterglow to flare energies in
both cases poses a puzzle for this last explanation as well.

\vspace{0.25in}

 This research was partially funded by a US-Israel BSF grant
and NASA ATP NNG05GF58G grant.  E. N. was supported at Caltech by a
senior research fellowship from the Sherman Fairchild Foundation. R.
S. is a Packard Fellow and an Alfred P. Sloan Research Fellow. We
thank Brian Cameron, Kevin Hurley, Jonathan Katz, Arieh Konigl and
Shri Kulkarni for helpful discussions.


\end{document}